\documentclass[12pt,a4paper%
]{article}
\usepackage{graphicx}
\usepackage{empheq}
\usepackage{amsmath}
\usepackage{amssymb}
\usepackage{amsfonts}
\usepackage{amsbsy}
\usepackage{bbm}
\usepackage{yfonts}
\usepackage{txfonts}
\usepackage{dsfont}
\usepackage{exscale}
\usepackage{pstricks}
\usepackage{eepic}
\usepackage{latexsym}
\usepackage{theorem}
\usepackage{tocloft}
\usepackage{enumerate}
\newcommand{\kommentar}[1]{}
\newcounter{com}

\newcommand{\beq}{\begin{eqnarray*}}
\newcommand{\eeq}{\end{eqnarray*}}
\newcommand{\beqn}{\begin{eqnarray}}
\newcommand{\eeqn}{\end{eqnarray}}

\newcommand{\N}{\varmathbb{N}}
\newcommand{\Z}{\varmathbb{Z}}
\newcommand{\Q}{\varmathbb{Q}}

\newcommand{\C}{\varmathbb{C}}
\renewcommand{\P}{\varmathbb{P}}

\newcommand{\Vir}{\mbox{\textgoth{V\!ir}}}
\newcommand{\U}{\mbox{{{\;{U}}\llap{$n\, \atop {} \,$}}}
}
\newcommand{\Uo}{\mbox{{{{\;{U}}\llap{$n\, \atop \circ \,$}}}}
}

\newcommand{\ket}[1]{
\left|
{#1}
\right\rangle 
}

\newcommand{\bra}[1]{
\left\langle 
{#1}
\right|
}

\newcommand{\lkor}[1]{
\left\langle 
\Psi_{h_1,k_1}(z_1)
\ldots
\Psi_{h_{{#1}},
k_{{#1}}}(z_{{#1}})
\right\rangle 
}

\renewcommand{\d}{\mathrm{d}}

\newcommand{\oin}[2]{ 
\underset{#1}{ \oint} 
\! \frac{\d {#2}}{2 \pi i} 
}

\newcommand{\pa}{\partial}

\newcommand{\s}{\lambda}
\newcommand{\fl}{\Gamma}

\newcommand{\eref}[1]{(\ref{#1})}

\newcommand{\ceref}[1]{(\mbox{\ref{#1}{\normalsize $_{\clubsuit}$}})}

\newcommand{\eeref}[1]{eq.\ (\ref{#1})}

\newcommand{\roc}{\rotatebox[origin=cc]{45}{c} }
\newcommand{\rob}{\rotatebox[origin=cc]{315}{b}\! }

\newtagform{spade}{(}{{\normalsize $_{\spadesuit}$})}
\newtagform{club}{(}{{\normalsize $_{\clubsuit}$})}
\newtagform{heart}{(}{{\normalsize $_{\heartsuit}$})}
\newtagform{diamond}{(}{{\normalsize $_{\diamondsuit}$})}

\newcommand{\q}{Q}
\newcommand{\A}{q}

\newcommand{\vacj}[2]{
\prod _{ {#1}=0 }^{n-1}
(c_{0}^{({#1})})^{{#2}_{#1}}
\left|
0
\right\rangle
}

\newcommand{\cwegiI}[5]{ 
(-)^{\overset{{#4} }{ \underset{ {#2}={#3} }
{\sum}}{#1}_{#2}} \;
\delta _{{#1}_{#3},1} 
\delta _{{#1}_{#4},1} 
\underset{
			{
			{#5}	=     0  \atop
			{#3}	\neq    {#5}  \neq  {#4}
			}
		}{\prod} 
(c_0^{({#5})})^{N_{{#5}}}
\left|
0
\right\rangle
}

\newcommand{\we}{we}
\newcommand{\We}{We}
\newcommand{\our}{our}
\newcommand{\Our}{Our}

\newcommand{\us}{us}

\newcommand{\thesis}{paper}

\begin{document}
\title{Indecomposable Representations in $\boldsymbol{\Z}_{\boldsymbol{n}} $ Symmetric $b,c $ Ghost Systems via Deformations of the Virasoro Field}
\author{Michael Flohr\footnote{email: {\tt flohr@th.physik.uni-bonn.de}}\hspace{0.2cm}  and Julia Voelskow\footnote{email: {\tt voelskow@th.physik.uni-bonn.de}}
\vspace{0.3cm} \\
Physikalisches Institut der Universit\"at Bonn\\
Nu\ss allee 12, 53115 Bonn, Germany \hspace{0.1cm}
}
\renewcommand{\thesection}{\Roman{section}}
\maketitle
\vspace{-10cm}
\begin{flushright}
BONN-TH-2006-01\\
hep-th/0602056
\end{flushright}
\vspace{+9cm}

\begin{abstract}
The Virasoro field associated to $b,c$ ghost systems with arbitrary integer spin $\s$ on an $n$-sheeted branched covering of the Riemann sphere is deformed. This leads to reducible but indecomposable representations, if the new Virasoro field acts on the space of states, enlarged by taking the tensor product over the different sheets of the surface. 
For $\s=1 $, proven LCFT structures are made explicit through this deformation. In the other cases, the existence of Jordan cells is ruled out in favour of a novel kind of indecomposable representations.  
\end{abstract}
	
\tableofcontents

\section{Introduction\label{Introduction}}

It has become clear during the last decade that so-called logarithmic
conformal field theories are not just a set of pathological special
cases but form a well-defined class of generalised conformal field theories with
numerous applications. Logarithmic conformal field theories share many
of the powerful properties with their ordinary conformal field theory
kinsmen, but feature, as their defining characteristic property, 
indecomposable representations. The best known example for a logarithmic
conformal field theory has central charge $c=-2$. If realised in terms of
the so-called $\theta\bar{\theta}$ ghost system of Zamolodchikov, this
theory cannot do without logarithms, as was first demonstrated in
\cite{Gurarie:1993xq} (see also \cite{Rozansky:1992td}).

In recent years, many logarithmic conformal field theories turned out to be
bona fide generalisations of rational conformal field theories. As such,
we would prefer to refer to such theories as `conformal field theories with
indecomposable representations'. The first example of a rational
logarithmic conformal field theory was constructed in \cite{Gaberdiel:1996np}. Actually, it also has central charge $c=-2$ and is again related to a
ghost system, the so-called symplectic fermions. 

Logarithmic conformal field theory is now a research field on its own. 
For some recent reviews, see
\cite{Flohr:2001zs,Flohr:2004ug,Gaberdiel:2001tr,Kawai:2002fu}. From a physics 
point of view, logarithmic conformal field theories appear naturally
in various models of statistical physics, for example in 
the theory of (multi)critical polymers \cite{Flohr:1995ea,Kausch:1995py,Rozansky:1992td},
percolation \cite{Flohr:2005ai,Watts:1996yh}, two-dimensional turbulence
\cite{Flohr:1996ik,RahimiTabar:1995nc,RahimiTabar:1996ki}, the quantum Hall effect
\cite{Gurarie:1997dw} and various critical (disordered) models
\cite{Caux:1995nm,Caux:1998sm,deGier:2003dg,Gurarie:1999yx,Maassarani:1996jn,Moghimi-Araghi:2004wg,Piroux:2004vd,Ruelle:2002jy}. Above that, there have been applications in Seiberg-Witten models \cite{Flohr:1998ew} and in     
string theory, in particular in the context of 
D-brane recoil \cite{Kogan:1995df,Kogan:1996zv,Lambert:2003zr,Periwal:1996pw}, and in pp-wave
backgrounds \cite{Bakas:2002qh}. Logarithmic vertex operator algebras
have recently attracted attention in mathematics 
\cite{Carqueville:2005nu,Huang:2002mx,Huang:2003za,Kleban:2002pf,Milas:2001bb,Miyamoto:2002ar}.  
Most examples that have been studied concern the $c=-2$ model,
but logarithmic conformal field
theories have also arisen in other contexts, see for example
\cite{Fjelstad:2002ei,Gaberdiel:2001ny,Lesage:2002ch,Rasmussen:2005hj}. 

In many known logarithmic conformal field theories, Ward identities with respect to the Virasoro modes $L_{1}, L_{0}$ become inhomogeneous, the action of the energy operator $L_0$ exhibits Jordan cells.
The above-mentioned $\theta \bar{\theta}$ system can be obtained naturally from a $b,c$  ghost system at $c=-2$ \cite{Knizhnik:1987xp} on a branched double cover of the Riemann sphere \cite{Krohn:2002gh}. 
This observation promts the question whether $c=-2$ ghosts on higher Riemann surfaces, or even ghosts with higher spin, can analogously be extended to logarithmic theories. 
\We{} adress this question using a deformation technique similar to the ansatz of \cite{Fjelstad:2002ei}.
Nevertheless, \we{}  make use of the properties of the algebra of modes rather than those of the operator product expansion (OPE). 
For $c=-2 $, the analogy turns out to be perfect for all genera, as \we{} show in the following. 
For higher spin ghost systems, \we{} rule out the pretty well-known Jordan-cell type representations. 
Instead, the action and the form of $L_0$ remains completely unaltered.
In \our{} framework, \we{} show  highest weight representations as subrepresentations and Jordan cells to be mutually exclusive. 
Nevertheless, \we{} construct examples of `new' indecomposable representations.
After formulating the general setup in \ref{setup},
\we{} derive constraints on the deformation, in section \ref{bedingung}.
Then, \we{} explore \our{} possibilities to obtain Jordan cells in \ref{funny}, and prove the existence of reducible but indecomposable representations in \ref{indecs}. 
In \ref{Examples}, \we{} pursue to discuss some examples:
First of all,  
\we{} generalise the results of simple ghosts in presence of $\Z_2$ twists to the $\Z_n$ symmetric case in \ref{c=-2}.
Following that in \ref{FreeFields}, the bilinear deformation, which is a well-defined theory if one sums over all sheet-labels, is considered.
\We{} find an interesting peculiarity -- a whole subalgebra can be rendered to remain undeformed.
After some remarks on multilinear ansatzes in \ref{multilinear}, in \ref{quadrilinearDeformation} a quadrilinear deformation is discussed. 
In this context, \we{} explicitly construct an example of an indecomposable representation.
The impact on the Ward identities is commented on in \ref{WardIdentities}. 
In \ref{Interpretation}, \we{} discuss the findings of \ref{setup},\ref{WardIdentities} and close with the summary \ref{summary} of \our{} results.

\paragraph{Conformal Field Theory (CFT)}
Local conformal symmetry in two dimensions implies an infinite number of conserved charges 
$$
Q_{\varepsilon} = \oin{}{z} 
\varepsilon (z) T(z)
$$
with $\varepsilon (z)$ an arbitrary function having a Laurent expansion in some vicinity of the origin.
The generator of conformal transformations of the metric, $T(z)$, is the holomorphic component of the two linearly independent components of the energy-momentum tensor.
One can reconstruct the conserved charges from the modes of the energy-momentum tensor, 
obtained by Laurent expansion $
L_n=\oin{}{z} 
\frac{1}{z^{n+1}} T(z)
$. This holomorphic component
$T(z)= 
\sum _n 
L_n z^{-n-2}
$ is also called Virasoro field, because its  
 modes satisfy the famous Virasoro algebra  
\begin{gather}
\left[ 
L_q, 
L_m 
\right] = 
(q-m) 	L_{q+m}
+ 
\frac{\hat{c}}{2}	{q+1 \choose 3}
\delta_{q+m,0},
	\label{Virasoroalg}
	\qquad
\left[ L_q, \hat{c} \right]=0 
\end{gather} for all $m,q \in \Z$.

\noindent
In conformal field theory, the notion of primary and quasi-primary fields is 
of particular importance.
A field  $\psi(w) = \sum _ n \psi _n w^{-n-h_{\psi }}$ is called primary, 
if for all $m,n \in \Z$ 
\begin{equation}
	\label{Wardprimary}
\left[ L_{m},\psi _n \right] 
		=
		(m(h-1)-n)\psi _{m+n},
	\end{equation} 
and it is called quasi-primary, if the relations 
\eqref{Wardprimary} are valid for all $n \in \Z$ and $\mbox{$|m| \leq 1$}$.
Applying primary fields  to the vacuum, $\underset{z\rightarrow 0}{\lim{}} \psi _h (z) \ket{0}$, one obtains \emph{highest weight states}, i.e. states satisfying
\begin{eqnarray}
	L_0\ket{h_{\psi }}& = &h_{\psi}\ket{h_{\psi }} \label{EnergyOperator} \\
 L_m \ket{h_{\psi }}& = & 0 \qquad \qquad \forall \; m>0 .
\end{eqnarray} 
Because of the first relation, $L_0$ is called energy operator.

The field-state isomorphism is a basic characteristic of all (logarithmic) conformal field theories. 
A module of the Virasoro algebra with highest weight vector $\ket{h_{\psi}}$ is the span of $\ket{h_{\psi}}$ and its linearly independent descendants 
$\left\{\mbox{$ L_{n_m} \ldots L_{n_1}\ket{h_{\psi}}:$}\right.$
$\left.\mbox{$0<n_1 \leq \ldots \leq n_m$}\right\}$ 
for all $n_i, m \in \N$. 
The number $\sum _{i=1}^{m} n_i $ is called the level of the \emph{descendant} of $\ket{h_{\psi}}$. 
The algebraic relations \eqref{Virasoroalg} lead, 
for certain values of the \emph{central charge} $c$, 
to further linear dependencies, which imply the existence of 
\emph{singular vectors}. 
These are highest weight states to submodules, 
which are orthogonal to the rest of the module and  can therefore be divided out. 
The submodule structure of Virasoro modules was investigated by Feigin and Fuchs \cite{Feigin:1983tg,Feigin:1988se}. 
They showed three different structures to occur: There are either no, one or infinitely many null vectors \cite{BPZ84}. 
The latter leads to the minimal models.
These are models, which are rational, i.e. contain only finitely many representations,  with respect to the Virasoro algebra. 
\paragraph{Logarithmic Conformal Field Theory (LCFT)}
It was  discovered in \cite{Gurarie:1993xq} that if the field content of a ghost system  at $\s=1$ (introduced later in the next paragraph), is enlarged by a twist $\mu$, correlation functions develop logarithmic divergencies.
This primary field $\mu$ has weight $h_{\mu}=-\frac{1}{8}$ and corresponds to an admissible representation in the Kac table. 
The proof of the correlator  $\left\langle\mu\mu\mu\mu\right\rangle$ to have logarithmic divergencies somewhere on the Riemann sphere simply used fundamental conformal field theory features: The associativity of the operator product expansion, the behaviour of primaries undergoing conformal transformations, in particular rotations, and monodromy requirements.
It was clarified by \cite{Flohr:2001tj,Gurarie:1993xq} that logarithmic divergencies correspond to \emph{reducible but indecomposable} representations. 
In contrast to the afore mentioned minimal models, 
in a logarithmic conformal field theory  submodules do not necessarily decouple from the rest of the module and thus cannot be divided out in all cases. 
The term `indecomposable' does not refer to irreducible representations for the rest of this \thesis. 
Scale invariance is not broken by the occurence of logarithms in the operator product expansion, 
because the appearing logarithms depend on the (scale invariant) crossing ratios. 

A characteristic of logarithmic conformal field theories is the existence of fields being neither primaries nor descendants. 
Usually, these are the logarithmic fields, also called `log-partners to some primary fields'. 
A further peculiarity is the existence of prelogarithmic fields. These are primary fields, whose operator product expansions yield logarithmic fields.
The field $\mu$ is an example of a prelogarithmic field.
To \our{} knowledge, it was conventional to  assume logarithmic conformal field theories to contain only representations which comprise `standard' Jordan cells, i.e.\  there is one primary field and $r-1$ quasi-primary `logarithmic partner fields', such that 
$$L_0\ket{h_{\psi },k} =h_{\psi}\ket{h_{\psi },k}+(1- \delta _{k,0})\ket{h_{\psi};k-1} \qquad k \in \left\{ 0, \ldots,r-1 \right\}.$$ 
Here, $r$ denotes the rank of the Jordan cell. 
The Ward identities for logarithmic partner fields containing $\mathcal{L}_{1},\mathcal{L}_{0}$ (the differential operator obtained by acting with $L_{x}$ on correlators, $ x \in \left\{ -1,0,1 \right\} $, is denoted by $\mathcal{L}_{x}$)  become inhomogeneous in this case
\begin{equation}%
	\label{Ward}		
	\mathcal{L}_x G(z_1,\ldots, z_n)	   = 
\sum_i ( z_i^{x+1}\partial _i+(x+1)z_i^{x}[h_{i}+\hat{\delta} _{h_i}]) 
	G(z_1,\ldots, z_n)=0 ,	
\end{equation}%
where $ G(z_1,\ldots, z_n)=\lkor{n}$. The conformal weight is denoted by the first index $h_i$, the second label $ k_i$ denotes the position of the logarithmic partner field in the Jordan cell, $\hat{\delta } _{h_i} \Psi_{(h_j;k_j)}=\delta _{ij}\Psi _{(h_j;k_j - 1)}$, and $\hat{\delta } _{h_i} \Psi_{(h_j;0)}=0.$

\paragraph{Ghost Systems
}
In space-time dimension two, the bosonisation theorem affords the existence of anticommuting fields with integer conformal spin and weight $\s$. 

The so-called $b,c$ ghosts enjoy the mode expansions
$$ b(z)=\underset{l \in \Z }{\sum} b_l z^{-l- \s } 
\qquad
c(z)=\underset{l \in \Z }{\sum} c_l z^{-l+ \s -1}, 
$$
and the modes satisfy the anticommutation relations
$$
\left\{ b_l ,c_{n} \right\} = \delta _{n+l,0}
.$$
The modes $b_{1-\s}, \ldots , b_{\s-1}$ are \textit{zero modes}, i.e. they annihilate the in- \emph{and} the outvacuum. Their conjugate modes, called $c $ zero modes from now on, are creators to both sides of a correlator.
For correlators on the complex plane, the naive calculation  $ \left\langle 0|1|0 \right\rangle
= \left\langle 0| \left\{ b_{m},c_{-m} \right\}|0 \right\rangle=
\left\langle 0| b_{m} c_{-m}\;\;|0\right\rangle +\left\langle 0|\;\; c_{-m}b_{m}|0 \right\rangle=0$ for $m \in \left\{ 1-\lambda,\ldots,\lambda -1 \right\}$ 
shows that the vacuum is orthogonal to itself; hence  
one needs the non-trivial outstate 
$\bra{c_{1-\s}  \ldots c_{\s -1}}$ to get a non-zero result. 
For higher genus other modes have to be included, see e.g. 
\cite{Alvarez-Gaume:1986mi,Alvarez-Gaume:1987vm,Green:1987sp}.
This is not of interest in \our{} context, because \we{} express higher genus correlators by correlators on the Riemann sphere with additional insertions of twists.
This is precisely possible when the non-trivial Riemann
surface possesses a global $\Z _n$ symmetry which allows
 to simultaneously diagonalise all monodromies around
 any of its branch points.
On the contrary, the non-trivial outstate for $g=0$ is the crucial fact allowing for \our{} deformation to work (and it is believed, that zero modes are linked to the existence of logarithmic conformal field theories in general).

Ghost systems can be realised by a free field construction, the coherent states $\mbox{:$e^{i\alpha \phi(z)} $:}$ corresponding to $b$ and $c$ have charges $\alpha=1$ and $\alpha=-1$ resp. \cite{dHoker807,Knizhnik:1987xp}. 
The ghost action (displayed here with locally flat metric), $ S_{\s}=\frac{1}{2\pi}\int_{\Sigma}\d ^{2} z (b\bar{\pa }c+ \bar{b}\pa \bar{c})$, specifies the Virasoro field
$ T_{\s}= (1-\s)(\pa b)c -\s b\pa c$ with central charge $c_{\s}=-2(6\s ^2 -6\s +1)$.  
It also reveals
that ghosts match the definition of $\s$-, $(1-\s)$-differentials resp. \cite{FK80,dHoker807}, and thus exist on any Riemann surface.

Recently, various applications were found for $\s =1 $ in condensed matter physics.  
This system has amazing features and has been investigated very thoroughly by now.
These `simple ghosts' are used to describe self-avoiding walks, 
the Haldane-Rezayi fractional quantum hall effect, percolation, the abelian sandpile etc. 
Furthermore, 
they are used to bosonise the supersymmetry partners of the reparametrisation ghosts.

Ghost system other than the $b,c$ series play an important role in physics as well. 
One can obtain the $\theta \bar{\theta}$ from the $ \s=1\;  b,c$ ghosts by formal integration of the $b$ field to a spin zero field. 
This requires the introduction of additional zero modes, $\theta^{+}_0$ as integration constant, and the conjugate $\theta^{-}_0 $ as an addend to the former $c$ field.
In presence of $\Z_2$ symmetry, it is possible to identify these new zero modes with the zero modes from the other sheet \cite{Krohn:2002gh}. 
Great success has been made by describing the abelian sandpile within this model
\cite{Mahieu:2001iv,Piroux:2004si,Piroux:2004bc,Piroux:2004vd,Ruelle:2002jy}.
It describes the surface of a 
sandpile in a simplified discretised manner.
The field $\mu$ is used to 
simulate borders.

One  obtains a system with weights $(1,1)$, the symplectic fermions, 
which has been considered by Kausch \cite{Kausch:2000fu}, if instead the $c$ field is formally differentiated to a spin 1 field.

\section{The Setting\label{setup}}
\paragraph{Ghost Systems with $\Z _n$ Symmetry}
In this \thesis, \we{}  investigate $b,c$ ghosts on algebraic curves with $\Z _n$ symmetry. %
Such curves
 can be parametrised by numbers 
$a_i \in \C ; L_i,n \in \N ,$ where $i=1,\ldots,L$, 
and can be expressed as the graph 
$$
\fl= \left\{ (y,z) 
:
\;y^n(z)=\overset{L}{\underset{i=1}{\prod}}(z-a_i)^{L_i} \right\}.
$$ 
Then $g= (n-1)(L-1)/2$ is the genus of the Riemann surface. 
If any $L_i \neq 1$, the curve is called $singular$. 
$\Z _n$ symmetry means that all branch points $a_i$ have the same ramification number, $L=nm$, and implies that the monodromies around the branch points are simultaneously diagonalisable. 
The monodromy group acts on meromorphic%
\footnote{ In the framework of conformal field theory, `meromorphic' fields are allowed to have conformal weights $h,\bar{h} \in \Z /2.$ } 
fields $\varphi(z)$ via the mappings 
\begin{equation}{}
\hat{\pi } _{a_i}
\varphi (z) 
	= \varphi (e^{2\pi i l }(z-a_i)+a_i) \qquad l \in \Z . 
\nonumber	
\end{equation}
\We{} denote the covering map by 
$z: \fl \rightarrow  \C \P ^1$ 
and choose %
local coordinates $y$ such 
that $ 	
	z(y)=a+y^n  
$ in the vicinity of a branch point $a$.

In \cite{Knizhnik:1987xp}, a $\rob \roc$ ghost system of 
arbitrary, fixed integer spin $\s $ is considered on each of the $n$ sheets of the  surface $\Gamma$. 

The introduction of a new basis of fields 
\begin{gather}%
b^{(k)}  =  \underset{l=0}{\overset{n-1}{\sum}}e^{-2\pi i l {\A }_k}\rob ^{(l)}, \qquad \qquad
c^{(k)}  =  \underset{l=0}{\overset{n-1}{\sum}}e^{2\pi i l {\A }_k}\roc ^{(l)},\\
 {\A }_k:    =   \frac{(k+ \s  (1-n))}{n}, 
\label{Ladung}%
\end{gather}%
 instead of the old 
$\rob ^{(l)},\roc ^{(l)}$, 
diagonalises the monodromy, namely, 
$\hat{\pi}_a b^{(l)}= e^{2\pi i {\A }_l}b^{(l)} $ 
and 
$\hat{\pi}_a c^{(l)}=e^{-2\pi i{\A }_l}c^{(l)} $%
.
Parenthetised upper indices are used to denote the $n$ sheets of the Riemann surface (from now on with diagonalised monodromy). 
The Virasoro field and the current have the same shape in the new and in the original fields. 
Thus, \we{} have 
$T^{(k)}(z)=
(\s -1)\mbox{:$ (\pa b^{(k)})c^{k}$:}(z)
-\s\mbox{:$ b^{(k)}c^{(k)}$:}(z)$ 
with modes $L_{n}^{(k)}.$ 
\We{} denote the `\emph{total} theory', which lives on all sheets, by the supersrcipt $tot$, e.g. $T^{tot}(z)$. This notation should not be confused with the \emph{full} theory, composed out of left- and right chiral parts.
The quantity ${\A }_k$ is the charge of the branch point with respect 
to the currents $j^{(k)}=   \mbox{:$b^{(k)}c^{(k)}$:}=\mbox{:$\rob^{(k)}\roc^{(k)}$:} $, which are 
 single-valued functions in the vicinity of the branch points 
\cite{Knizhnik:1987xp}.  
In this work, it was concluded that the branch point can be represented by an insertion of a primary field with charge ${\A }_k$, hence, that $\Z_n$ symmetric ghost theories can be represented by a free field construction. 
In this formalism, the bosonised version of the twist would read 
$V_{\boldsymbol{\A }}(a)=\mbox{:$e^{i \sum_k {\A }_k \varphi _k}$:} (a)=:\mbox{:$e^{i {\boldsymbol{ {\A } \varphi}}}$:}(a)%
\label{twist}. 
$
Because the conformal weight of a coherent state is quadratic in its charge,
$
h= \sum_k h_k=  \sum_k (\frac{1}{2}{\A }_k^2+(\s -\frac{1}{2}){\A }_k)
$, the coherent states with charges  
${\A }_k $ and  $2{\q }-{\A }_k=:{\A }_k^*$ have the same conformal weight, where  $c= 1-24\q ^2= -2(6\s ^{2} -6\s +1)$. 

It was noticed in \cite{Kogan:1997fd} that even for  $  {\A }_k={\q }$ there  are two primary fields with the same weight $h_{\q}$. 
The second one is not a coherent state, but a so-called puncture operator. 

\paragraph{Two Conflicting Views of Simple Ghosts in Presence of $\Z_2$ Twists}
The twist field for $\Z_2$ symmetry coincides for $\s=1$ with Gurarie's field $\mu$ at $h_{\mu}=-\frac{1}{8}$, which is a prelogarithmic field.
Thus, some of the modules of this theory have to be indecomposable.
These indecomposable structures are somehow hidden in \cite{Knizhnik:1987xp}, 
 because the points of twist insertions had been fixed to describe the 
geometry of specific non-singular Riemann surfaces. 
When two branch points come close to each other, 
e.g. when the OPE is inserted, 
the represented Riemann surface gets a new puncture while the genus is reduced by one. 

Two striking inconsistencies emerge from the results of Gurarie and Knizhnik. 
\begin{enumerate}%
\item 
Variation of the bosonised action
with respect to the metric in the so-called twist field formalism, 
 yields an energy-momentum tensor, which acts perfectly diagonalisable on the space of states, which is in contradiction to the results of Gurarie. 
If twists are present in the theory, a modification of the Virasoro field, and, consequently, of the bosonised action, is needed to unveil the logarithmic structure.
\item 
Knizhnik's bosonisation formul\ae{} do not mirror the facts  proved by Gurarie.
By calculating the OPE of two of the bosonised twist fields, 
one immediately sees that they, 
being constructed from ordinary coherent states, 
do $not$ lead to logarithms nor logarithmic partner fields
\begin{eqnarray}%
(z-z')^{-\boldsymbol{\A }\boldsymbol{\A } ^*}V_{\boldsymbol{\A }}(z) V_{\boldsymbol{\A }^*}(z')= 
\frac{\mathbbm{1}}{z-z'}+ j(z)+T(z)(z-z') + reg(z-z')
. 
\end{eqnarray}%
This requires a redefinition of the bosonisation formul\ae{}. 
\end{enumerate}%
This \thesis{} is concerned with the resolution of the first inconsistency. 
For the second inconsistency, 
a mathematically well-defined bosonisation scheme of the 
(prelogarithmic) twist field first had to  be established.
First attempts in this direction have been made via so-called puncture operators, 
see \cite{Kogan:1997fd}. 
\paragraph{\Our{} Knowledge on Ghost Systems with Higher Spin or Higher Twists}
As shown in \cite{Eholzer:1998se} the twisted sectors of different ghost systems are connected via a spectral flow, which suggests common features of these models. 
Nevertheless, it is not known whether ghost systems with $\s>1$ are genuinely subtheories of logarithmic ones. 
However, this question needs an approach completely different from that of 
\cite{Flohr:2001zs,Gaberdiel:2001tr,Gurarie:1993xq} which is mainly based on the representation theory of the Virasoro algebra. 
The $c=-2$ ghost system with $\Z _{n > 2} $ symmetry are not accessible to the analysis of Gurarie for the following reasons: 
In \cite{Kausch:1995py,Kausch:2000fu,Gaberdiel:1998ps} it
was shown that for all twists the $\mathcal{W}(2,3)$ algebra is contained in the extended symmetry algebra.
There are  indications
that the symmetry algebra is an extension of a  $ \mathcal{W}(A_{n-1})$-algebra. In \cite{Wang:1997wj,Wang:1998bt} the modules of $\mathcal{W}(2,3)$ were classified.
There are no null vectors which would yield differential equations allowing a reasoning similar to Gurarie's \cite{Gurarie:1993xq}.
However, in \cite{Kausch:2000fu} a co-multiplication formula of  \cite{Gaberdiel:1996kx} was exploited to derive a differential equation on twist-correlators in the $(1,1)$-symplectic fermion model with  $\mathcal{W}(2,3)$ as symmetry algebra.
 
The ghost systems with $\s>1 $ do not contain any null vectors wrt the Virasoro algebra alone.
It is merely conjectured that these ghost systems are rational. 
These maximally extended algebras cannot be in the set of the already classified $\mathcal{W}$-algebras \cite{Flohr:1994sm,Krohn:2002gh} and  are unknown, as are the corresponding singular vectors.

\section{The Deformation}
Recall that \we{} use parenthetised upper indices as labels for the diagonalised sheets. 
For improved lucidity, only the letters $r,s,t$  label sheets from now on.
By the same token, \we{} use the letters $w,x,y$ in lower indices to indicate conformal weight of the Virasoro and the deformation modes, for the contributing $b$,$c$ and $\beta$ modes, the characters $l,k,m,h,i$ are in use as well. 
\subsection{Derivation of the Constraints \label{bedingung}}

To find a way to do a similar analysis as Gurarie did,
\we{} investigate whether indecomposable structures could arise, if a deformed Virasoro field acts on enlarged spaces of states.
This techniqe was used in \cite{Fjelstad:2002ei}, and gave -- in the case of the $c=-2$ model -- rise to a term, which resembles the extra term in the Virasoro field of the $\theta \bar{\theta}$ system.
But in contrast to the approach used in \cite{Fjelstad:2002ei}, \we{} demand the deformation to consist out of `known' modes, i.e. to have a mode expansion with values in  
the universal envelope of the algebra of $b_x^{(r)},c_{y}^{(s)}$, with $r,s \in \Z_n$ and $x,y \in \Z$. It is denoted by $\U$, 
its subsets, the universal envelope of the algebra of $b_x^{(r)}$ or $c_x^{(r)}$ modes, with $r \in \Z_n, x \in \Z$, are called $\U _b$ and $\U _c$ resp.
The universal envelopes of the algebras of zero modes (as above, but $x,y \in \left\{ 1-\s, \ldots ,\s -1  \right\}$) are termed $\Uo$, $\Uo _b$ and $\Uo _c$ resp.  
\We{} suppress the obvious dependence on $\s $.
The deformed $ {\tilde{L}}_{x}^{(s)}$ \textit{and} the original $ L^{(s)}_{x}$ modes of the energy momentum tensor must fulfil Virasoro commutation relations to the same central charge. 
This entails 
\begin{equation}
U^{(r)}_{x+y} \delta_{r,s}
       =	
         	[L^{(r)}_x,U^{(s)}_y] 
        -	[L^{(s)}_y,U^{(r)}_x]
        +	[U^{(r)}_x,U^{(s)}_y]
	\label{Virasorodifferenz}
\end{equation}
for the deformation modes 
$ 
\tilde{L}_{x}^{(s)}-L^{(s)}_{x} = U^{(s)}_{x}.
$
For the beginning, \we{} assume $ \left[ U^{(r)}_x,U^{(s)}_y \right]=0$ for all $x,y \in \Z$ and all $r,s \in \Z _n$. 
This simplification covers the cases where  $\beta ^{(r)} _{x} \in \U _b$. 
This simplification is justified further in the next subsection
and it is an expedient ansatz, because the ground states are obtained by application of elements of $\Uo _c $ on the vacuum. 
We put up 
$U^{(r)}_x:= 
\sum_{l}     	P(\s  ,x,l,r)    \beta^{(r)}_{l}b^{(r)}_{x-l}$
and attempt to find a functional   $	P(\s  ,x,l,r)$ 
so that the new modes $\beta^{(r)}_{l}$ 
can be identified with elements of $\U _b$. 
In the next section \ref{funny}, \we{} briefly comment whether restraints are put on the representations by \our{} simplification. 
In the appendix, \we{} discuss whether there actually are deformations with  $ \left[ U^{(r)}_x,U^{(s)}_y \right]\neq 0$.
Continuing with \our{} simplified ansatz, the single consistency condition splits into independent ones. 
The conditions \eqref{Virasorodifferenz} then read

\begin{gather}
(x-y)\underset{l}{\sum}
	P(\s  ,x+y,l,s)
		\beta^{(s)}_{l}b^{(s)}_{x+y-l}
			\delta_{r,s}
=
\label{constraints} \\
= \underset{k,l}{\sum}P(\s  ,y,l,s)(\s  x-k)b^{(r)}_k 
\left\lbrace
 	 c^{(r)}_{x-k},\beta^{(s)}_{l} 
\right\rbrace
 b^{(s)}_{y-l} 						
 \tag*{	{\normalsize$\clubsuit$}} 		\\ 
 + \underset{l}{\sum}
 P(\s  ,y,l,s)(\s  x-x+l-y)\beta^{(s)}_{l}b^{(s)}_{y+x-l} 
 \delta_{r,s}	
 \tag*{{\normalsize $\spadesuit$}}	\\
-
\underset{k,m}{\sum}(\s  y-k)b^{(s)}_k P(\s  ,x,m,r) 
\left\lbrace
  c^{(s)}_{y-k},\beta^{(r)}_{m} 
\right\rbrace
b^{(r)}_{x-m} 
\tag*{{\normalsize $\clubsuit $}}	\\ 
- \underset{m}{\sum}P(\s  ,x,m,r)(\s  y-y+m-x)
\beta^{(r)}_{m}b^{(r)}_{x+y-m} \delta_{r,s}
 \tag*{{\normalsize $\spadesuit$}}	
.
\end{gather}
Rows marked by different suits contain linearly independent vectors of the universal envelope of the $b^{(r)}_{m},c^{(s)}_{l},\beta^{(t)}_{o}$ algebra. 
Now \we{} investigate what algebraic relations the $\beta^{(t)}_{o}$ and the $b^{(r)}_{m}, c^{(s)}_{l}$ modes might fulfil.
\noindent
\usetagform{spade}%
Comparing coefficients yields 
\begin{gather}
P(\s  ,y,l,s)(\s  x-x+l-y) - P(\s  ,x,l,s)(\s  y-y+l-x)
 \\ \nonumber
 =
(x-y)P(\s  ,x+y,l,s) \qquad
 \forall l,x,y  .
\end{gather}
Insertion of  $y=0 $ turns this recurrence relation into an explicit one, 
\usetagform{default}%
\begin{equation}
\fbox{$ 	
 P(\s  ,0,l,s)	
\cdot	( (\s  -1)	\cdot x+l ) 
= l \cdot P(\s  ,x,l,s)
\tag{{\Large $\spadesuit$}}
\label{virasoroj}.
$
}
\end{equation} 
One finds that the solution factorises into one part for $l$, $\s  $, $s$ fixed 
and an up to now unspecified part which could be a functional depending on $l$, $\s  $, $s$.
The recurrence relation can thus be solved, 
provided $ P(\s  ,x,l,s)$ satisfies the condition
\begin{equation}
P(\s  ,x,l,s)=A_{\s  ,l,s} \cdot ((\s  -1)\cdot x+l) 
\tag{${}^{ {\Large \spadesuit }}_{ { \Large\spadesuit}}$}
\label{virasolj}.
\end{equation}%
\usetagform{spade}%
Note that for $\s  =1$ this is independent of $x$, and allows for $any$ dependence on $l$.  
In this case, there is a solution $independent$ of $l$. 
If then only $l=0$ is allowed to contribute, i.e. $A_{1 , l,s} =A_{1 , 0,s}\delta _{l,0}$, the deformation of $L_0^{(s)}$ contains only zero modes of conformal weight 0.
For $\Z _2$ twists, this yields the deformation suggested in \cite{Fjelstad:2002ei}. 
As long as $\s  >1$, $L_0$ cannot contain terms, which consist out of modes with conformal \mbox{weight 0} only, due to the additional factor of $l=0$.
Then, \eeref{virasoroj} reveals that the deformation terms have to have exactly the same shape as the original Virasoro modes, with the $c$ modes replaced by $\beta $'s. 
One might be tempted to arrange them in the same manner  as  the modes constituting the original Virasoro modes. 
This however leads to a recurrence relation which does not decouple and thus seems not to be solvable in an acceptable amount of time and effort.
Instead, \we{} deliberately ordered them the other way round -- and got a relation which is easily solved.
Relation (\ref{virasoroj}) constitutes one of the main results of this \thesis.
\usetagform{club}%
Imposing the solution \eref{virasolj} yields
\usetagform{default}%
\begin{equation}
\fbox{
\mbox{
$ \displaystyle
A_{\s  ,l,s}
	\left\lbrace  
		c^{(r)}_{m},\beta^{(s)}_{l} 
	\right\rbrace 				
+A_{\s  ,m,r}
	\left\lbrace  
		c^{(s)}_{l},\beta^{(r)}_{m} 
	\right\rbrace				
 =0.
$
}}
\tag{{\Large $\clubsuit $}}
\label{gammaj}	
\end{equation}%
\subsection[Restraints on the Representation Theory for $\s >1$]{Restraints on the Representation Theory  for $\boldsymbol{\s }\boldsymbol{ >}\boldsymbol{1}$ \label{funny}}
The Virasoro constraints for the particular choice of modes 
\begin{gather}
	[L^{(r)}_x,U^{(s)}_0] 
-	[L^{(s)}_0,U^{(r)}_x]
\overset{!}{=}	
x \cdot U^{(s)}_x \delta_{r,s}.\label{special}
\end{gather}
reveal that Jordan cells are  impossible for $\s >1$ if
\begin{itemize}
	\item only additive deformations of the Virasoro field by elements of $\U$ are taken into account, and
	\item the zero mode of the Virasoro field is required to act as energy operator.
\end{itemize}
The following utilises the primarity of the $b$'s and $c$'s wrt to the \textit{old} 
 Virasoro modes repeatedly. (Note that the improvement terms do not need to respect  primarity, in particular, of the $c$'s.) Thus, \eqref{special} is equivalent to 
\begin{equation}
[L^{(r)}_x,U^{(s)}_0] \overset{!}{=}0 \; \; \forall  x 
.  
	\label{ }
\end{equation}
In case only the total new Virasoro algebra contains an energy operator,
\begin{gather}
\underset{r,s}{\sum}	[L^{(r)}_x,U^{(s)}_0] 
-
\underset{r,s}{\sum}	[L^{(s)}_0,U^{(r)}_x]
\overset{!}{=}	
x \underset{s}{\sum}U^{(s)}_x 
\end{gather}
implies at least that 
\begin{equation}
	[L^{(r)}_x,U^{tot}_0]=
	\underset{s}{\sum}	[L^{(r)}_x,U^{(s)}_0] = 0\qquad \forall x .
\label{KeineJordanzelle}
\end{equation}
In \our{} framework, these sets of equations generically require $U^{(s)}_0$ to be zero. 
Direct calculation yields  
 that exemptions are possible only for $\s  =1$. 
The relation
$$
[L^{(s)}_x,b^{(r)}_0]=((0)x-0)	b^{(r)}_{x}	 \delta_{r,s} =0,
$$  
affords a non-vanishing  $U^{(s)}_0$ for $\beta \in \Uo _b$. 
The well known $c=-2$ LCFT deformation is such an example.%
\footnote{
This also implies that we cannot find other additive deformations of the Virasoro field which have Jordan cells with rank greater than two, because the deformation of $L^{(s)}_0$ then at least has to contain $b^{(s)}_0$ as an overall factor.
Otherwise the individual Virasoro algebras are either trivially deformed, 
or do not commute.}
There are no further non-trivial and well-defined solutions to eq. $\eqref{KeineJordanzelle}$.
Nevertheless, because the maximally extended algebra is not known, \we{} cannot decide whether $U^{(s)}_0 $ has to commute with its  generators. 
Therefore \we{} do $not$ conclude that Jordan cells are forbidden in a larger algebra, 
although there can be no Jordan cells with respect to $L^{tot}_0$, 
as long as it is required to measure the conformal weight correctly.
\subsection{New Indecomposable Structures\label{indecs}}
Let \us{} introduce an antilinear involution $^{\ddagger}$ on $\U$ 
which maps the subsets $\U_b $ to $\U_c$ and vice versa via the prescription 
\begin{alignat}{6}
	\nonumber
&U&=&\sum_l a_l & b^{ (r_{ 1l }) } _{i_{1l}}  
\ldots 
b^{ ( r_{ml} ) } _{ i_{ml} } 
&\in \U _b &\text{ with }  a_l \in \C^{ \times}, \text{ then} \\
& U^{\ddagger}&= &
\sum_l \frac{1}{a_l}&
c^{(r_{ml})}_{-i_{ml}} \ldots  c^{(r_{1l})}_{-i_{1l}} &\in \U_c. &%
\nonumber	\label{}
\end{alignat}
This offers a convenient way to show how indecomposable structures can arise and how these are linked to the non-trivial vacuum structure and the zero modes. 
If a deformation mode $U_y^{(r)}, \;|y|< \frac{\s (\s -1)n}{2}, $ contains an element of $\Uo _b$ as an addend, 
$(U_y^{(r)})^{\ddagger} \ket{0}= \sum _i \ket{\psi}_i$ is a 
sum of states, which comprises a ground state, e.g. $ \ket{ \psi }_j$. 
Therefore, $(U_y^{(r)})\sum _i \ket{ \psi }_i = m \ket{0}+ $descendants, 
where the positive integer $m $ is the number of pure zero mode products in $U_y^{(r)}$. 
This shows that the irreducible 
(because all deformation modes vanish on it) vacuum module 
is contained in the modules of those groundstates which can be obtained by application of an 
$U_y^{(r)})^{\ddagger}$ on $\ket{0}$. 

Under certain circumstances, the module of $\ket{\prod_{r \in \Z_n}c_{1-s}^{(r)}\ldots c_{s-1}^{(r)}}$ contains $\ket{0}$ as a highest weight vector of a submodule, which cannot be divided out, because it is not orthogonal to the rest. 
Then,  it necessarily has to contain at least one further submodule, 
because the deformation can only kill the $c $ zero modes, 
and its conformal weight differs from zero, such that the action of a single deformation cannot suffice. 
An example is discussed at the end of section \ref{FreeFields}.
Note that the indecomposable structure becomes obvious only in the action of the ${\tilde{L}}_y ^{(s)}$ with $ y \neq 0$.
To \our{} knowledge, this is a completely new kind of indecomposable structure, at least, in CFT or physics applications. 
It was discussed in \cite{Rohsiepe:1996qj} that for the `minimal' series $c=c_{1,q}= 1-6\frac{(q-1)^2}{q}$ so-called staggered modules appear.
The action of $L_0$ on an ancestor of a staggered module yields a \emph{descendant} in another  module. 
It was remarked that more complicated structures could occur. 
In \our{} case, the action of the other Virasoro modes may yield `ordinary' descendants plus \emph{highest weight vectors} in a module different from that the Virasoro modes acted on.
In contrast to an LCFT with the usual Jordan cells, the Hilbert space in \our{} deformed theories with $\s >1$ is graded with respect to the energy operators. 
However, in contrast to usual CFT, the representations of the ghost charge do not coincide with the deformed Virasoro modules, because the currents do not commute with the deformation.

\section{Realisations of the Deformation\label{Examples}} 
\subsection[The ${c}{=}{-}{2}$ Extended Theory with Additional ${\Z} _{{n}}$ Symmetry]{The $\boldsymbol{c}\boldsymbol{=}\boldsymbol{-}\boldsymbol{2}$ Extended Theory with Additional $\boldsymbol{\Z} _{\boldsymbol{n}}$ Symmetry%
\label{c=-2}}
\usetagform{club}%
\We{} now investigate the ansatz $A_{1,l,s}=A_{1,0,s}\delta_{l,0}$, or more precisely  
$\; U^{(r)}_x	= 
\beta ^{(r)} 	b ^{(r)} _{x}
\;$.
The action of the new Virasoro modes on the various ground states may result in additional conditions if the matrix representation of $ {\tilde{L}} _0 ^{tot}$ is to contain Jordan cells.
\We{} denote the ground states as 
$\vacj{r}{N}$	
with 
$
	N_r	\;	\in \; \{ 0,\; 1 \}
.$ 
The ordering prescription is that 
$c^{(r)} $ resides on the left of $c^{(t)}$ if $ 0 \leq r  < t \leq n-1$. \\ \noindent 
With $\; 
\beta ^{(r)} := 
\sum _{s=0}^{n-1}
M_{sr}	b ^{(s)} _{0} 
$, the deformation
acts as
$\; U ^{(r)} _0 
	\vacj{t}{N}	=
$\\
$$= 
-\overset{n-1}{\underset{s=r+1}{\sum}}
	M_{sr}
\cwegiI{N}{t'}{r}{s}{t}
 +
\overset{r}{\underset{s=0}{\sum}}
	M_{sr}  
\cwegiI{N}{t'}{r}{s}{t}
.$$
In the last step, \we{} `rearranged' signs by raising the summation index in the exponential 
to avoid a space-consuming case discrimination. 
The summation is understood to give the same result if upper and lower bound are interchanged.

Thus, if one denotes 
$ \mathcal{P}_{rs}:= \mathrm{span}
	\{
		\left|
			\psi
		\right\rangle 
		 =\vacj{t}{N}: N_r,\: N_s \neq 0
	\}
$
it follows that
\begin{equation}
\fbox{
$M_{sr}	\neq	M_{rs} \; 
\Rightarrow 
{\tilde{L} _0 } ^{tot}
\left|
	\psi
\right\rangle
	\notin
 \mathcal{P}_{rs}.
$
}
\end{equation} 
This assures rank-two Jordan cells in the action of suitable fields, regardless of the number of sheets.
All ground states are highest weight states with respect to the deformed Virasoro modes, ${\tilde{L}}_x ^{(s)} \vacj{t}{N}=0 $ for all $x>0$, because only $c_0$ modes contribute. 
 This shows that the extension of the energy-momentum tensor leads to logarithmic divergencies in OPEs. 
The results above are in perfect concordance with the results of \cite{Kausch:2000fu}. 
Considering weight $(1,1)$ symplectic fermions, Kausch could prove that for all ramification numbers, 
the functional dependence of the four-point function 
$\langle 
\mu_{q_1}(\infty) 
\mu_{q_2}(1) 
\mu_{q_3}(x) 
\mu_{q_4}(0) 
\rangle$ with $  q_i \in \Q
$ 
on the crossing ratio aquires at most one logarithm as a factor. 
This exactly happens if either $q_3 +q_1 \in \N $ or $q_3 +q_2 \in \N $ or $q_3 +q_4 \in \N $, the other sums are fixed by charge balance. 
In other words, the twists have to be dual to each other.

\subsection{A Deformation by Free Fields \label{FreeFields}}
\usetagform{club}%
 Apart from their nilpotent action, the new Virasoro zero modes have to act as  energy operators
$\tilde{L}_0^{(r)}\ket{h}=h\ket{h}+\cdots$.
This implies that the conformal weights on the different sheets of $\tilde{L}_0^{(r)}$ vanish separately. 
This implies $\beta^{(r)}$ to have non-vanishing conformal weights only on the $r$th sheet. 
The following deformation destroys the property of the ${\tilde{L}_0^{(s)}}$ to be energy operators, but as $L_0^{tot}$ remains undeformed, the sum over the different theories is well-defined. 
The choice   
$
\beta ^{(s)} _{l} =:
\overset{n-1}{\underset{r=0}{\sum}}
M_{rs} b ^{(r)} _{l}
$
leads to the requirements

\begin{equation}
\fbox{$ 
\displaystyle
	M_{rs}
A_{\s ,l,s}
+
	M_{sr}
A_{\s ,-l,r}
\overset{!}{=}
0
\label{mist}
$}
\end{equation}
\usetagform{default}
by virtue of \eqref{gammaj}.

\usetagform{default}%
If one allows $M_{rr}$ to be non-zero on the $r$th sheet,  $A_{\s  ,-l,r}$ is fixed as well, 
\mbox{$A_{\s  ,l,r}=A_{\s  ,-l,r}$}.  
A term of the form $M_{rs}A_{\s  ,l,s}$ occurs with a prefactor of \mbox{$((\s  -1)x+l)$} in  $ U^{(s)}_x$ for $every \; x$ and some $l$ if any $L^{(s)}_y$ is to be deformed. 
The only possibility to deform $ T^{(s)} (z)$ without changing a distinct mode, say $ L^{(s)} _w$, is to demand 
$
A_{\s  ,l,s}=A_{\s  ,(1-\s  )w,s} \delta _{(1-\s  )w,l}
,$ 
because then 
$ U^{(s)}_x$ reads 
$\sum_s
	M_{rs}
		A_{\s  ,(1-\s  )w,s}
		((\s  -1)x+(1-\s  )w)
b^{(r)}_{(1-\s  )w}
b^{(s)}_{x-(1-\s  )w},$ 
which obviously vanishes if and only if $x=w$, 
provided $A_{\s  ,(1-\s  )w,s}$ 
and any of the $M_{rs}$ are non-zero. 
Due to eq. \ceref{mist},  $ U^{(r)}_{-x}$ 
vanishes, but
$
 U^{(r)}_{x}
$ 
is forced to be non-zero. 
This implies in particular that if on any sheet, say $s$, $L^{(s)}_y$ is to be deformed,
there is another sheet, 
for instance called $r$, 
on which the translation operator $L^{(r)}_{-1}$ is deformed.

In the well-defined \textit{total} theory, one 
can choose a whole subalgebra (spanned by $L_{\pm y}^{tot}, L_0 ^{tot}$) of \Vir{} for one $y$ to remain undeformed. 
Because of this, one can even enforce `non-logarithmic' Ward identities. 
Because of \eeref{mist}, 
 a non-zero $A_{\s ,l,r}M_{sr}$ implies a non-zero $A_{\s ,-l,s}M_{rs}$. 
 For a term in an arbitrary deformation mode $U^{tot}_{-y}$, one can find exactly one second term with the same mode content, by inversion of the second index in $A_{\s ,l, s}$ at $y$.
This implies $A_{\s ,l, s}M_{rs} $ and  $A_{\s ,y-l,r}M_{sr}$ to be prefactors of terms with the same mode content in $U^{tot}_y$.
To abolish one specific $U_y ^{tot}$, fix $A_{\s ,l, s}M_{rs} $ for one tuple $(l,s,r)$ and adjust other prefactors recursively, depending on the $y$ chosen. 
This leads to a series of constraints
\begin{gather}
\left.
\begin{array}{r@{\; = \; }l}
	  -((\s -1-i)y+l)
		A_{\s  ,l-iy,s}
		M_{rs} 
&
		((\s +i)y-l)
		A_{\s  ,(i+1)y-l,r}
		M_{sr} 
 \\
- ((\s -1-i)y-l)
		A_{\s  ,-l-iy,r}
		M_{sr} 
&
	((\s +i)y+l)
		A_{\s  ,(i+1)y+l,s}
		M_{rs} 
	\end{array}
\right\rbrace	
				\forall i \in \N. 
\label{seriesofconstraints}
\end{gather}
By \ceref{mist}, every adjustment evokes a new term.  
This could be cancelled by further adjustments, as denoted in the figure below.
It depicts which prefactors have to be equalled -- those joined by vertical arrows -- and which mutually imply each other -- those connected by horizontal arrows.
\\
$$\includegraphics[scale=0.5]{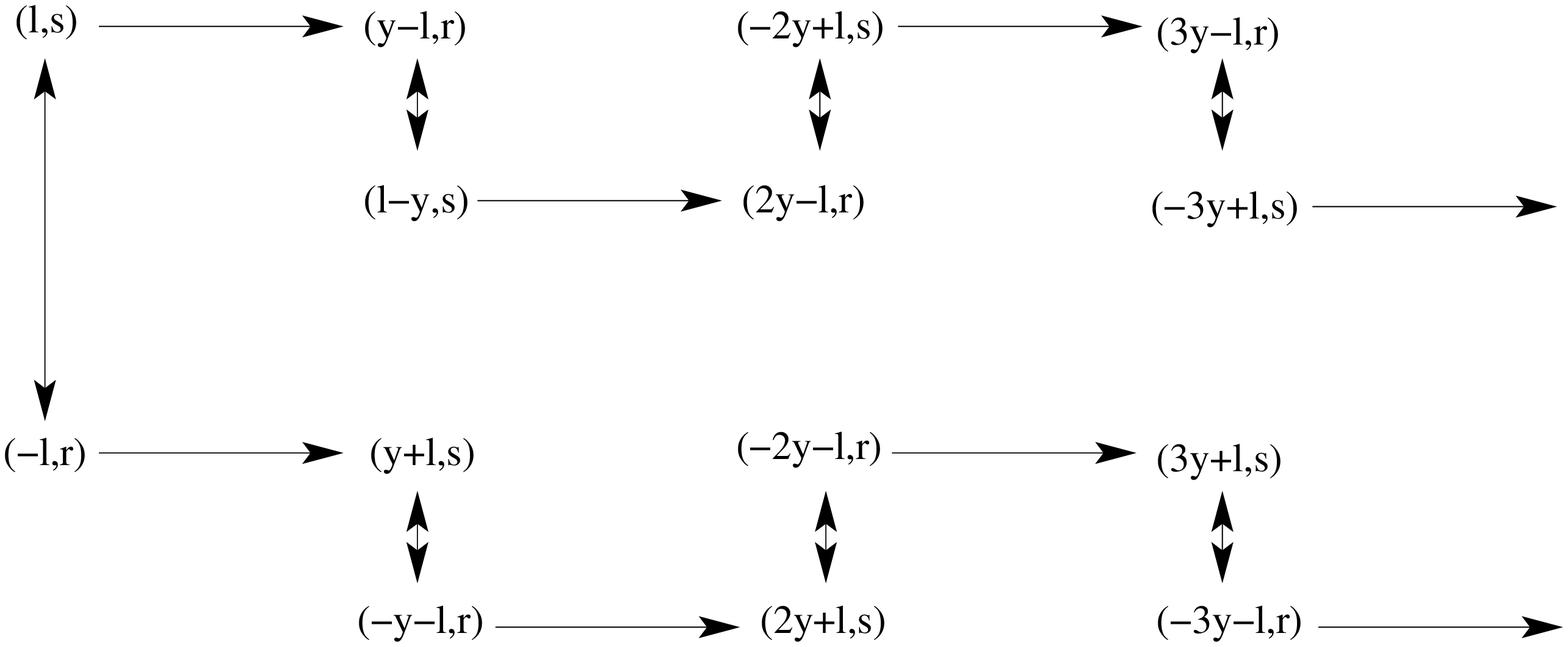} $$
Generically, if these conditions are to be met, 
infinitely many terms have to contribute to the individual deformation modes. 
This extends to terms which do not contain zero modes any more and are mappings from excited states with negative ghost charge to such with positive ghost charge. 
However, if the `initial' value of $l$ is zero, 
there is only one condition in the first place.
Furthermore, the sum terminates at 
$(\s -1)y=i$. 
It suffices to adjust all $A_{\s ,(i' +1)y,r}M_{sr}$ with $|i'|< (\s -1)|y|$ to satisfy \eqref{seriesofconstraints}. 

The choice of $l$ to be zero implies that the equations emanate from each other by permuting their sheet-labels.
Only the prefactor $A_{\s,y\s,r} $ has to vanish for all sheets $r$.
However, eq. \ceref{mist} then implies that this equation is still valid when the sign of $y$ is flipped. 
Thus, with a vanishing $U^{tot}_y$,  $U^{tot}_{-y}$ also vanishes. 
Choosing $y \neq 1$ leads to an impact on the Ward identities as discussed in \ref{WardIdentities}.
A similar behaviour of an improved $\Z_2$ twisted $c=-26$ theory was noted in \cite{Krohn:2002gh}.

The action of the deformed Virasoro modes yields descendants and states with  ghost charge diminished by two, which are of the same level wrt the total theory.

When the ramification number is even, $\ket{0}$ is a singular vector in the module of 
$
\ket{\prod_{r \in \Z_n}
	c_{1-\s}^{(r)}
		\ldots 
	c_{\s-1}^{(r)}}
\propto 
\overset{\frac{n}{2}-1}{\underset{t=0}{\prod}} 
\big(
	U_{-\lambda +1}^{(2t+1)}
	U_{\lambda}^{(2t+1)}
	\overset{-1}{
	\underset{m=-2\lambda +3}{
	\prod}}
	\overset{2\lambda -3}{
	\underset{1}{
	\prod}}
	U^{(2t+1)}_{m}  
\big)^{\ddagger}
\ket{0}$, 
because e.g. 
\begin{samepage}
$$
\ket{0}=
\overset{\frac{n}{2}-1}{\underset{t=0}{\prod}} 
b_0^{(2t)} 
b_{-\lambda+1}^{(2t+1)}
b_{\lambda -1}^{(2t)}b_1^{(2t+1)}
\overset{ {\lambda -2}}{\underset{ { i=0}}{\prod}}
\overset{ { 1}}{\underset{ {i=-\lambda +1}}{\prod}}
b_{i}^{(2t)}
b_{i+1}^{(2t+1)}
\ket{\prod_{r \in \Z_n}
	c_{1-\s}^{(r)}
		\ldots 
	c_{\s-1}^{(r)}} \propto
$$
$$
\propto 
U_{-\lambda +1}^{(2t+1)}
U_{\lambda}^{(2t+1)}
\overset{-1}{
\underset{m=-2\lambda +3}{
\prod}}
\overset{2\lambda -3}{
\underset{1}{
\prod}}
U^{(2t+1)}_{m}
\ket{\prod_{r \in \Z_n}
	c_{1-\s}^{(r)}
		\ldots 
	c_{\s-1}^{(r)}}
.$$
	\end{samepage}
As an example consider a $ \s =2 $ ghost system with $\Z_2 $ symmetry. 
There is only one summand in $(U_{-1}^{(1)})^{2}U_2^{(1)}$ which is in $\Uo _b$ and it is characterised by $l=0, l=-1$ in the first two factors and $l=1$ in the last.  
Thus, if $A_{2,0,1}, A_{2,-1,1}$ and $A_{2,1,1}$ are non-zero, this combination of deformation modes maps $\ket{c^{(0)}_{-1}c^{(0)}_{0}c^{(0)}_{1}c^{(1)}_{-1}c^{(1)}_{0}c^{(1)}_{1}}$ to $4A_{2,0,1}A_{2,-1,1}A_{2,1,1}\ket{0}\propto \ket{0}$. 
Note that this indecomposable structure may exist even if $y=1$ \eqref{seriesofconstraints}, i.e. even, if $U_1^{tot}\overset{!}{=}0$ which implies that the Ward identities used in CFT are valid. 
This is an example of an indecomposable structure, for which the highest weight vector of a submodule is \emph{not} orthogonal to the ancestor of the module.
Thus, even on hyperelliptic Riemann surfaces, such an indecomposable structure does not necessarily have filtration length  two \cite{Rohsiepe:1996qj}.

This should work analogously for other deformations, if the number of modes in the deformation term is a divisor of the ramification number.

The deformation of reparametrisation ghosts on hyperelliptic surfaces found in \cite{Flohr:2003tc,Krohn:2002gh} is not an example of this type of deformation, although it is constructed similarly. 
But this model is somewhat special, 
because the different Virasoro algebras do not commute with each other. 
The identification of the additional zero modes was made `cross-over', 
such that a certain zero mode would act as an annihilator on one, and a creator on the other sheet. 
It seems impossible to generalise this to higher twists. \\

\subsection{A Bosonic Deformation with Adjustable Nilpotence Index \label{multilinear}}

The choice 
$$
U ^{(s)} _y:=  
\underset{l}{\sum}P(\s  ,y,l,s) A^{(s)} _0 b^{(s)}_{l}b^{(s)}_{y-l} 
=
\underset{l}{\sum}P(\s  ,y,l,s) A^{(s)} _0 B^{(s)}_{y}
,$$
with $ A^{(s)}_0$ being composed out of modes whose sheetwise conformal weight adds up to zero $by \; pairs$, leads to the requirements
\begin{gather}
\mbox{
\fbox{$ 
\displaystyle
A _{\s  ,l,s}=- A _{\s  ,-l,s}
$}
}
\label{antisymmetry}
\end{gather}
for all sheet labels $s$ and all integers $l$.
\We{} set 
$$\displaystyle A^{(s)}_0:=\:\:\underset{h>0}{\sum}M(s,h)
\underset{ t\neq s}{\prod}
\big(
b^{(t)} _{-h}	b^{(t)} _{h}
\big) ^{N_{h,t}^{(s)}}
$$%
and find the conditions
\begin{equation}
\fbox{
\mbox{
$A_{\s  ,h,r}M(r,l) N^{(r)}_{l,s}
=
A_{\s  ,l,s}M(s,h)N^{(s)}_{h,r}.$
}
}
\end{equation}
which have to hold for all sheet labels $r,s \in \left\{ 0, \ldots ,n-1 \right\}$ and for all integers $l, h$.

The nilpotency index of the deformation\footnote{The nilpotency index of the deformation is the least number $q$ such that $(U^{(s)}_y)^q=0$} can be lowered without the need to restrict to a certain number of different contributing zero modes.
This could be interesting if one does not want quadratic contributions of the deformation  in normal ordered products of the new Virasoro modes, e.g.\  if one tries to keep generators of an extension to a $\mathcal{W}$-algebra.

\subsection{A Quadrilinear Deformation \label{quadrilinearDeformation}}

Of particular simplicity is the similar ansatz
$$	
A^{(s)} _0= \underset{m>0,t}{\sum}
M^{ts}_m	b^{(t)} _{-m}		b^{(t)} _{m}	
,$$
which yields
\begin{equation}
\fbox{$
\displaystyle
A_{\s  ,l,s}	M^{rs}_m=
A_{\s  ,m,r}	M^{sr}_l
 \qquad A_{\s  ,l,s}= - A_{\s  ,-l,s}
 \label{vermischtes}
$}
\end{equation} 
for all $l,m \in \left\{ 0, \cdots , \s-1 \right\}$ and all sheet labels $ r,s \in \left\{ 0, \ldots ,n-1 \right\}.$
\We{} observe that if the $A_{\s  ,l,s}
$ are chosen to be the same on all sheets, $M^{rs}_m$ is a symmetric matrix for all $m>0$. 
Furthermore, if any $M^{rs}_m, A_{\s  ,m,r}$ are non-zero, then $A_{\s  ,l,s}=0 \Leftrightarrow M^{sr}_l=0$.
This implies that some translation operator gets affected by the deformation. 
The impact of this on the Ward identities is investigated in the next section.
The above, in particular eq. \eqref{vermischtes} implies that we have $ 
\frac{n^2(j-1)^2+n(j-1)}{2}$ independent deformation `directions'. 
(We can choose one independent coefficient for any pair of tuples 
$\left( r,s,m,l \right) \neq \left(s,r,l,m  \right),\: $
the entries $r,s,l,m$ in the same ranges as above,
and one for each 
$ \left( r,s,m,l \right) = \left(s,r,l,m  \right)$.)
By \eeref{antisymmetry} the value $l=0$ cannot be chosen. 
The set of indices $l$ to be summed over then had to be infinite. 
This had as a consequence that summands in the deformation term  altered the ghost  charge of excited states, 
while vanishing on the corresponding ground states. 
Above, \we{} excluded such implicitly by restricting to $A_{\s  ,l,s} = 0 $ if $|l| \geq \s $.

\subsection{New Logarithmic Ward Identities \label{WardIdentities}}

One may now ask whether there are highest weight states  or states corresponding to quasi-primary fields except for the vacuum. 
\We{} restrict \our{} examination to states which originate from the action of $b^{r}_x$ and $c^{s}_y$ on $\big|0\big\rangle$. 
Thus \we{} do not look at the representations $\big|h\big\rangle$, $h \notin \Z$, e.~g. the represetations corresponding to the twist fields.
With respect to the total old Virasoro algebra, 
only the  $\mathfrak{sl}(2,\C )$-invariant vacuum is a highest weight state. 
It is easy to see that every state, 
which satisfies the condition that
\begin{gather}
N_{l_s}=1
\Rightarrow
N_{(l+1)_s}=1
\label{Viralt}
\end{gather}
for all of  its  occupation numbers,    
is quasi-primary with respect to the individual old Virasoro modes $\Vir ^{(s)} _{old}$.

\noindent
To investigate which states $\ket{\psi}$ are quasi-primary with respect to the new Virasoro modes, ${\tilde{L}_1}\ket{\psi}=0$, 
it suffices to  look for states, which satisfy \eeref{Viralt}
and remain invariant under the deformation mode $U_1^{(s)}$. 
Because the deformation lowers the mode content, 
both the Virasoro and the deformation term 
with conformal weight $-1$ have to vanish separately on these states. 
Evidently, $U_1^{(s)}$ vanishes on states $\ket{\psi } $ which cannot be written as $(U_1^{(s)})^{\ddagger}\ket{\psi ^{\prime} } $.

In case of a quadri- or multilinear deformation this happens if $U_1^{(s)}$ does not contain those suitable pairs of $c$ modes, 
$c^{(s)}_{-l}$ and $ c^{(s)}_{l-1} $.
Thus, a state is quasi-primary with respect to the new Virasoro modes, if it contains only modes 
with conformal weights $\leq 0$ on the relevant sheet 
and  satisfies condition \eqref{Viralt}. 

Now consider a corresponding $U^{(s)}_{-1}$. 
By the same reasoning as before, 
it vanishes on all states which do not simultaneously contain $c^{(s)}_{-l}$ and $ c^{(s)}_{l+1} $. 
But while the product of modes $c^{(s)}_{0}c^{(s)}_1$ vanishes under the action of $U^{(s)}_{1}$, 
it could give a non-zero result under the action of $U^{(s)}_{-1}$, 
even  in suitable \Vir$_{old}$ quasi-primary combinations with other modes. 
On the contrary, there are no inhomogeneities possible in the Ward identity corresponding to 
$\widetilde{\mathcal{L}}_0$. 
Furthermore, there could exist quasi-primary fields satisfying homogeneous differential equations derived from $ \widetilde{\mathcal{L}}_{-1}$. 
These necessarily satisfy homogeneous differential equations with respect to 
$ \widetilde{\mathcal{L}}_{1}$. 
The inverse of the last statement is \textit{not} true. 
There \textit{are} quasi-primary fields, i.e.\   which enjoy homogeneous Ward identity involving  
$\widetilde{\mathcal{L}}_1$ but have inhomogeneities wrt $\widetilde{\mathcal{L}}_{-1} $. 
An example is a field corresponding to 
$
\big|\hat{(s)}\big\rangle
\otimes \big| c_{0}^{(s)}c_{1}^{(s)}\big\rangle $, with $\big|{\tiny \hat{(s)}}\big\rangle$ containing at least one pair $c_{-i}^{(r)}c_{i}^{(r)}$ with $ r \neq s,$ and $ A_{\s,l ,s }M^{rs}_i \neq 0 $ for at least one $l$.
Even more astonishing, for the bilinear deformation, one encounters indecomposable respresentations, while CFT Ward identities are valid.

To summarise, the indecomposable structures one runs across here 
have the property that $L_0 $ is unaltered 
(and thus blind with respect to the indecomposable structure), 
and suitable states are annihilated by ${\tilde{L}}_1$, 
but not by ${\tilde{L}}_{-1}$ at the same time. 
In particular, the indecomposable structure becomes visible in the action of ${\tilde{L}}_{-1}$, or $U_{-1}$, respectively. 
This is a completely different and new kind of indecomposable representation (to LCFT), since the ordinary Jordan-cell-type representations are distinguished by the feature that $L_{-1}$ remains undeformed.
This shows that some of the conditions in \cite{Flohr:2001tj
} were indeed necessary, 
to generalise the Ward identities in such a simple and elegant way to \eeref{Ward}.  

\subsection{Interpretation of the Results
\label{Interpretation}}
One could interpret the vanishing of the deformation of $L_0$ and the modification of the action of $L_{-1}$ on some quasi-primary states for $\s > 1$ as follows: 
In theories compactified on the torus, 
the generators of the global conformal group still act as differential operators
$\mathcal{L}_i , i \in \left\{ -1,0,1 \right\}$ 
corresponding to translations, scalings and special conformal transformations on quasi-primary fields. 
But the transformation to the new coordinates for radial quantisation maps scalings, translations 
and special conformal transformations not separately onto themselves, 
but only the whole group. 
A rotation in the old coordinates, generated by $L_0$, corresponds to a translation in the new and 
$\mathcal{L}_{1}$ generates rotations instead. 
It is worth noting in this context that for ghost theories with integral spin $\s >1$,  
we only know applications -- namely, the bosonic string -- 
where the worldsheet is naturally compactified on a torus (in light-cone coordinates), 
in contrast to condensed matter applications of $c=-2$, 
which naturally live on the complex plane. 

One could speculate that indeed these theories may have applications only in cases, 
where in some sense the indecomposable structure manifests itself in the action on the space of states of the generator of rotations, 
whereas the translation operator acts diagonalisable.

\section{Summary\label{summary}}
Indecomposable representations are predicted by \cite{Gurarie:1993xq,Kausch:2000fu} for the $\Z _n$ symmetric $c=-2$ ghost system, but not seen in the action of the Virasoro field.
To resolve this problem, \we{} introduce additional `deformation' terms
$
	U^{(s)}_y:=
	{\sum_{l \in \Z }}
		P(\s  ,y,l,s)
		\beta^{(s)}_{l}	b^{(s)}_{y-l}
$. 
\Our{} deformation ansatz works for all ghost systems of integer spin $\s$, not only for $c=-2$, i.~e. $\s =1$.
The new modes $ \beta^{(s)}_{l}$ were  taken from the universal enveloping algebra $\U _b$ of thealgebra of modes $b_m^{(r)}, m \in \Z, r \in \Z _n$.
\We{} derived constraint equations and found the general shape of the solution
\begin{gather}
 P(\s  ,0,l,s)	
\cdot	( (\s  -1)	\cdot x+l ) 
= l \cdot P(\s  ,x,l,s)
\tag{\ref{virasoroj}}
\\
A_{\s  ,l,s}
	\left\lbrace  
		c^{(r)}_{m},\beta^{(s)}_{l} 
	\right\rbrace 				
+A_{\s  ,m,r}
	\left\lbrace  
		c^{(s)}_{l},\beta^{(r)}_{m} 
	\right\rbrace				
 =
	0 \qquad  A_{\s  ,l,s}:= P(\s  ,0,l,s).
\tag{\ref{gammaj}}	
 \end{gather}
These equations were also used to prove that if one demands the Virasoro zero mode $L_0$ to remain an energy operator under the deformation and $\s \neq 1$, Jordan cells are impossible for \our{} deformations, the deformation zero mode has to vanish.
This statement was made using the primarity of the $b$'s and $c$'s wrt the old Virasoro modes and therefore is made only for $\beta \in \U$.
The new Virasoro modes cease to commute with the currents.
Thus, their respective representations do not coincide. 
\We{} showed indecomposable representations to exist: 
The modes $L_{n\neq 0}$ yield the usual descendants plus highest weight vectors from other modules. 
Under certain circumstances, this even happens though the whole global conformal group remains unaltered.
In section \ref{c=-2}, \we{} constructed a logarithmic extension of the $c=-2$ model  for arbitrary twists by adding 
$	U^{(r)}_m := \mbox{:$ \beta^{(r)} b^{(r)}_{m} $:} $ 
to the Virasoro modes  
as a straightforward generalisation of the construction in \cite{Fjelstad:2002ei}. 
\We{} explicitly investigated the action on the space of states of the ansatz  
$
\beta ^{(r)} := 
\sum^{n-1}_{s=0}
M_{sr}	b ^{(s)} _{0}. 
$
The action of the modified energy operators ${\tilde{L}}^{(r)}_0$, $ r \in \left\{ 0,\ldots,n-1 \right\}$, on the space of ground states shows the desired rank 2 Jordan cells. 

\We{} showed that the simple ghost system ($\s =1$) is the only fermionic ghost system that allows for this construction. 
Rank 2 Jordan cells for energy operators ${\tilde{L}_0}$ could furthermore be shown to be possible if and only if  
$\beta \in \Uo _b$.
\We{} scrutinised a bilinear deformation, for which ${\tilde{L}_0^{tot}}=L_0^{tot} $ is an energy operator, and constructed an explicit example of an indecomposable representation.
Without affecting this indecomposable structure, one can adjust a whole subalgebra ${\tilde{L}}_0^{tot},{\tilde{L}}_{\pm y}^{tot}$ of $\Vir $ to be undeformed.
With multi- and quadrilinear deformations, \we{} found examples of deformations which are well-defined (${\tilde{L}_0^{(r)}}=L_0^{(r)} \forall r$) even on the individual sheets.
For the quadrilinear deformation, there are quasi-primary states $\prod^{\s-1}_{i=1}|c^{(s)}_{i}\rangle |C ^{\widehat{(s)}}\rangle  $ which   
 vanish under $U^{(s)}_{ 1}$, but not under $U^{(s)}_{- 1}$.
Here, $|C ^{\widehat{(s)}}\rangle  $ denotes an arbitrary state with insertions of $c$ zero modes from sheets other than $s$.
Thus, the global conformal Ward identities have  to be altered further for the general case. 
Hence, the specific conditions, under which the LCFT Ward identities could originally be derived, 
were necessary indeed.
In the appendix the most general ansatz is commented on.

\section{Outlook\label{Outlook}}
A very interesting question arising from \our{} considerations is
whether  the deformations found above are consistent with different extensions of the Virasoro algebra. 
This could be the maximally extended $\mathcal{W}$-algebra, or a superextension.
It is unclear whether for \mbox{$\s >1$} there could be extended algebras which admit Jordan cells for the zero modes of other generators.

Furthermore, it remains to be investigated whether the indecomposable structures found for higher spin $\s$, which lack Jordan cells, lead to logarithmic singularities in correlators. 
A possibility to verify  this would be to find a modified current $\jmath^{(r)}(z)$ which is consistent with 
$\widetilde{T}^{(r)}(z)= 
	\frac{1}{2}\mbox{$:\jmath^{(r)}\jmath^{(r)}:$}(z) 
+
	(\frac{1}{2}-\s)\partial \jmath^{(r)}(z)$. 
If one were able to show that $\jmath_0$ had Jordan cells, 
the theory would clearly exhibit logarithmic divergencies.
Also, if the zero mode of this current were to commute with all Virasoro modes, 
$\jmath _0$ would respect the ${\tilde{L}}_0$-grading of the Hilbert space.
This would imply that $L_0$ maps states to sums of states which reside in different superselection sectors. 
This also would be a proof for the theory to be logarithmic.
Further examination of the anomaly of the current $j$ and, if found, the deformed current  $\jmath $  seems to be necessary.

Another interesting question would be to explore the antiperiodic sector, and, in particular, the Ramond sector of the fermionic `ghost' systems with half-integer spin $\s$.
It is as yet unclear whether the twist remains primary with respect to the modified Virasoro modes. 
Because it has no mode decomposition except in the bosonic language, which proved not to be  trustworthy by \cite{Gurarie:1993xq}, 
one would have to use the twisted Borcherds identity of \cite{Eholzer:1998se} to decide that. 
Particularly exciting would be 
a bosonisation scheme of the twists which is compatible with the OPE.
Then, one should investigate the indecomposable structure with respect to
 the maximally extended chiral symmetry algebra. For example, it is
 known that the $c=-2$ LCFT model is rational with respect to
 the triplet ${\cal W}$-algebra ${\cal W}(2,3,3,3)$. As we have seen, this
 model is given by the $\Z _2$ symmetric case of the spin $\s =(1,0)$
 ghost system. However, there exist consistent CFTs for the $\Z _n$
 symmetric case for all $n$. The spectrum of the twist fields of these theories 
 can be found in the Kac tables of ${\cal W}$-minimal models of the
 ${\cal W}A_{n-1}$ series at $c=-2$. Interestingly, the value $c=-2$ for the
 central charge appears in the respective minimal series of all these models.
 At precisely this value, the corresponding ${\cal W}$-algebras collapse to one 
 common object, the so-called unifying ${\cal W}$-algebra 
 \cite{Blumenhagen:1994ik}. Recently, the representation theory of the
 $\Z _n$ symmetric case was investigated in \cite{Abe0503}, and it
 would be interesting to see how this relates to the minimal models of the 
 ${\cal W}A_{n-1}$ series. 
 
 Finally, nothing is known about whether the deformations of $\Z _n$ 
 symmetric $(\lambda,1-\lambda)$ ghost systems, $\lambda >1$, which we 
 constructed in this work, possess extended chiral symmetries. This is in so 
 far an important issue, as these models clearly are not rational with respect 
 to the Virasoro algebra alone. The corresponding twist fields do not even 
 appear as degenerate representations in the Kac table at the central charges
 $c_\lambda$. Thus, if our models, which show indecomposable structures, 
 are  rational models, this must be entirely due to the
 extended chiral symmetry. Otherwise, they
 might yield examples of non-rational logarithmic conformal field theories.
 In any case, these models presumably will yield a new class of logarithmic
 theories, as their indecomposable structure is of an as yet unknown type.
\clearpage
 \vspace{0.7cm}
	 \noindent
	 \begin{samepage}
		{\bf Acknowledgements.}\\   
	 \noindent
Julia Voelskow would like to thank David Heilmann and Jan Schumacher for their hospitality. 
She is very grateful to Nils Carqueville for his careful reading of this document. 
Michael Flohr wishes to express his appreciation to Marco Krohn for stimulating discussions and comments.
The research of Michael Flohr is supported by the European Union Network
 HPRN-CT-2002-00325 (EUCLID). 
	 \end{samepage}
\appendix
	\section{General Considerations on the Deformation Modes\label{generalisation}}
\subsection[Constraints for Deformations 
which Contain $c$ Modes]{Constraints for Deformations 
which Contain $\boldsymbol{c}$ Modes}
The question arises whether all possible deformation terms in $\U$ are in $\U_b$. \We{} display \our{} incomplete reasoning, which could be used to rule out specific deformations, though \we{} could not answer this question in general yet.
As soon as $\beta$ is regarded as an element of $\U$, one has to demand that $ [U^{(r)}_x,U^{(s)}_y]$ vanishes separately.
Two normal ordered products of modes are obviously linearly independent if they comprise different numbers of factors.
It seems to be impossible to find modes $\beta_x ^{(r)} \in \U \setminus \U _b$  which match the requisite conditions. 
Counting factors reveals that
\begin{itemize}
	\item $\mbox{:$[U^{(r)}_x,U^{(s)}_y]$:} $ has to vanish for all $r,s,x,y$
	\item $\left([U^{(r)}_x,U^{(s)}_y]-\mbox{:$[U^{(r)}_x,U^{(s)}_y]$:}  \right)  $ has to vanish \textit{or} to contain the same number of $b^{(t)}, c^{(t)}$ modes  as $U^{(r)}_x$ and $U^{(s)}_y$  for all  $r,s,x,y,t$. 
		In the second case, 
this contributes to the rest of \eqref{Virasorodifferenz}%
.	
\end{itemize}
The non-trivial second case (of the second item) and the first item imply that 
exactly one half of the factors of $U^{(r)}_x$ are conjugate modes to the ones of  $U^{(s)}_y$. 
Furthermore, the commutator has to contain the same mode twice, if this has no conjugate, and at least two different modes twice, if one of them has a conjugate in the product. 
Besides, one expects that the zero mode content would be less than zero.
(Nevertheless, nilpotent products of $N$ $b$ modes and $N $ $c$ modes, e.g. $b_2 b_3 c_{-1}c_{-4}$, are possible).
Sadly, \we{} cannot rule out the $\beta _x ^{(r)}$ to reside in $\U \setminus \U _b$.
Then, by normal ordering, the analogue to eq. \eqref{constraints} would be a whole hierarchy of equations with different mode contents.

If $\beta$ is a linear combination of $b$'s and $c$'s, the commutator vanishes. 
Because a linear combination $\beta^{(r)}_l:=\sum _{s}M_{sr}c^{(s)}_l$ would otherwise alter the central charge, $M_{ss}=0$ for all $s$. Therefore $U^{(r)}_{x+y}\delta_{r,s } \equiv 0$ and the old and new Virasoro modes are the same. 

\subsection[Some Remarks on the Constituents of the Field 
$ {\beta}$ for ${\lambda} {>} {1}$]{Some Remarks on the Constituents of the Field 
$ \boldsymbol{\beta}$ for $\boldsymbol{\lambda} \boldsymbol{>} \boldsymbol{1}$}
The following considerations might prove useful for someone who wants to construct a specific deformation.
The most general form of $\beta$ allowing for $\left[ L^{(s)}_0,U^{(r)}_x \right]= -x\delta_{r,s}$ is 
$$\beta_l^{(r)}= 
\underset{i=0}{\overset{n-2}{\sum}}
\underset{ { I , r \notin I}}{\sum}
\underset{j \in I }{\prod}
W
_{ I }
(
l,r)A_0^{(j)}(l,r)B_l^{(r)}
$$
\We{} now use $B_l^{(s)},A_0^{(k)}(l,s) $ to denote products of $b$ modes from the $s$th and $k$th sheet resp. (Polynomials can be obtained by linear combination.)
The symbol ${\sum_{I}}$ denotes the sum over all possible subsets $I$ of $\left\{ 0, \ldots , n-1\right\}$ with $i$ elements.
Insertion into \eqref{gammaj}
reveals the following three equations  
(we  shuffle any normalisation into the factor 
$W
_{ I}(
m,s)$ without loss of generality) 
for all $ m,l \in \Z, r,s \in \Z_n$  
$$
A^{(j)}_0(m,s)=A^{(j)}_{0}(l,r),
$$
if the index $j$ is taken from the same set $I$ on both sides,
$$
B^{(s)}_{-m}=\left( \widehat{b_{-m}^{(s)}}A^{(s)}_{0} \right),
$$
$$
\left( A_{\s,m,s}
W
_{ I }
(
m,s)
-
A_{\s,l,r}
W
_{I' }
(
l,r) \right)\delta_{ I \!\setminus r, I' \!\setminus s }=0.
$$ 
These equations put severe constraints onto product ansatzes. 
Any deformation can thus be written as a sum of atomic deformations satisfying 
$$
U_{x}^{(\sigma (a))}=
\underset{m}{\sum}
( (\s -1)m+x)
	A_{\s ,m, a} 
	\underset{i \in I, i \neq a \in I}{\prod}
	A_0^{(\sigma (i))}
	B_m^{(\sigma (a))} 
	b^{(\sigma (a))}_{x-m}
	.$$%
(By the previous equation, the factors 
$W_{ I }(m,s)$
can be absorbed into $ A_{\s,m,s}$. The permutation $\sigma $ of $i$ elements accounts for the fact that fixing a deformation on one sheet fixes all others contributing to the atomic deformation.)

Furthermore, if the deformation term is to have physically sensible `hermiticity' properties (because of the non-trivial vacuum structure, hermiticity is merely a formal concept), the $A_0^{(s)}$ have to resemble those of the quadri- or multilinear ansatz \emph{or} to consist of summands of the form 
$
\underset{i \in J}{\prod}
\Big( 
	d_i d^{\prime}_{-\underset{k \in J}{\sum}k}
\pm 
	d_{-i} d^{\prime}_{\underset{k \in J}{\sum}k}
\Big)
$
where $J$ is an arbitrary proper subset of $\Z$ and 
$d, d^{\prime } \in 
\left\{ b^{(s)},c^{(t)}:s,t\in \Z \right\} $.
Above that, this fixes $A_{\s,m,s}=-A_{\s,-m,s}$, and \we{} have 
$(U_{x}^{(s)})^{\dagger}
	=(\pm)^{j}
	\frac{-A_{\s,-m,s}}{A_{\s,m,s}} 
	U_{-x}^{(s)}
$, where $j$ denotes the number of elements of $J$. 
Thus, hermicity can also depend on the number of contributing sheets in contrast to the cases analysed in the main part of this \thesis.
The deformation characterised by 
$P(1,x,m,s)=A_{1,0,s}\delta_{m,0}$, 
complies with 
$(U_{x}^{(s)})^{\dagger}=-U_{-x}^{(s)}$. 
The other deformations (the bi-, quadri- and multilinear ones), 
satisfy 
$(A_{0}^{(s)})^{\dagger}=A_{0}^{(s)}$. 
If $A_{\s,m,s}=\pm A_{\s,-m,s}$, and thus 
$P(\s,x,m,s)= \mp P(\s,-x,-m,s)$,
for all
$s \in \Z _n , m \in \Z$, 
then this entails $(U_{x}^{(s)})^{\dagger}=\pm U_{-x}^{(s)}$.
\newcommand{\etalchar}[1]{$^{#1}$}

\end{document}